\documentclass[12pt]{article}

\usepackage{epsf}
\usepackage{latexsym,amssymb,euscript}
\usepackage[dvips]{graphicx}
\usepackage{amsmath}

\begin{document}

\begin{center}
{\Large \textbf{Dual formulations of non-abelian spin models: local 
representation and low-temperature asymptotics}}

\vspace*{0.6cm}
\textbf{O.~Borisenko\footnote{email: oleg@bitp.kiev.ua}, \
V.~Kushnir\footnote{email: vkushnir@i.kiev.ua}} \\

\vspace*{0.3cm}
{\large \textit{N.N.Bogolyubov Institute for Theoretical
Physics, National Academy of Sciences of Ukraine, 03143 Kiev, Ukraine}}
\end{center}

\begin{abstract}
Non-abelian lattice spin models with symmetry group $SU(N)$ or $U(N)$ 
can be formulated in terms of link variables which are subject to the 
Bianchi constraints. Using this representation we derive exact and 
local dual formulation for the partition function of such models on 
a cubic lattice in arbitrary dimension $D$. Locality means that the dual 
action is given by a sum over some subset of hypercubes of the dual 
lattice and the interaction between dual variables ranges over 
one given hypercube. Dual variables are in general discrete-valued and 
live on $(D-2)$-cell of the dual lattice, in close analogy with the $XY$ 
model. We use our construction to study in details the dual of $SU(2)$ 
principal chiral model in two dimensions. We give dual expressions also 
for two-point correlation function in arbitrary representation and for 
the free energy of defects. Leading terms of the 
asymptotic expansion of the dual Boltzmann factor are computed and it 
is proven that at low temperatures it converges to a certain Gaussian 
distribution uniformly in all fluctuations of dual variables. This result 
enables us to define the semiclassical limit of the dual formulation 
and to determine an analog of the vortex--spin-wave representation for 
the partition function. Such representation is used to extract leading 
perturbative contribution to the correlation function which shows 
power-like decay at weak coupling. We also present some analytical 
evidences that the low-temperature limit of the dual formulation is completely 
described by $ISO(2)$-like approximation of $SU(2)$ matrix elements. 
\end{abstract}

\section{Introduction}

Two-dimensional ($2D$) lattice spin models, besides being interesting on
their own right as models of ferromagnets can be viewed as mathematically
well defined scheme of the nonperturbative regularization of quantum continuum
field theories like $SU(N)$ principal chiral model or $O(N)$ nonlinear sigma
model. Being a discrete theory lattice models can be simulated on computers
by Monte-Carlo method, and this is nowadays the most important tool
of obtaining physical results near the continuum limit. Among major analytical
methods one could mention the strong coupling expansion and the perturbation
theory (PT). The PT is essentially the only analytical tool which provides
systematic expansion of different physical quantities in the region of weak
coupling, i.e. in the region relevant for the construction of the continuum
limit. However, at the best PT is only applicable for studying short-distance
quantities and the most important and interesting phenomena, like the mass gap
generation cannot be described in its frameworks.
When attempting to go beyond PT in the region of weak bare
coupling, one runs into various mathematical problems, the most important being
an absence of any analytical control over exponentially small contributions
and, hence the absence of any reliable method to study long-distance physics 
where nonperturbative effects dominate even if they are exponentially suppressed.
These facts impelled people to look for different, though equivalent
representations which would allow to study the long-distance physics. One of
such popular and promising representations is known as a dual formulation
and is based on a certain non-classical change of variables.
This formulation deals with dual lattice and appears to be very
fruitfull for abelian models \cite{savit}. Namely, within such formulation
1) have been obtained various important analytical results on
the long-distance dynamics of abelian spin fields and 2) the topological
structure of the vacuum is transparent and can be more easily studied.

In the context relevant to this paper we would like to mention
the dual of the abelian $XY$ model \cite{savit} which has been used to prove
the existence of a soft phase at low temperatures with power-like decay of
the correlation function in two-dimensional $XY$ model \cite{rigbkt}.
In this case the dual of the $XY$ model is a local theory for certain
discrete variables. The conventional dual transformations \cite{pfeiffer}
for non-abelian models are perfectly well defined in mathematical terms and also
lead to a local dual theory for integers which label irreducible
representations of the symmetry group. Nevertheless, these transformations
are not complete if one compares to the abelian case. First of all,
the resulting dual variables are not independent
but are subject to some constraints known as triangular conditions, and as such
they cannot really be associated with elements of a dual lattice (of course,
introduction of the dual lattice is not the crucial point but rather
matter of convenience; e.g., in the abelian case one can work with
dual variables on the original lattice). It is to say, however that
it is not really clear what are true independent degrees of freedom
of the duals of non-abelian models. Secondly, such formulation is so
mathematically involved that today  it is even far from obvious
if it can be useful for any kind of the study of the model, except for 
the strong coupling expansion. In particular, the positivity of the Boltzmann 
weight remains unclear (see discussion in \cite{pfeiffer}). Even the precise 
definition of what can be termed the dual Boltzmann weight
is not really obvious, so the possibility of numerical simulations
is at present rather uncertain. Most importantly, however is that it is not
clear how one could proceed in an analytical study of the model at
low-temperatures\footnote{This is, of course personal opinion of the authors.
After some attempts to understand low-temperature properties of the dual
Gibbs measure we cannot say whether it possesses any kind of well defined
asymptotic expansion at large $\beta$.}. 

On the other hand, there exists representation of $SU(N)$ and $U(N)$ spin models 
in terms of link variables \cite{linkrepr}, and this representation can be formulated
directly on the dual lattice. It is a first goal of the present paper to use
the link representation for the derivation of an exact dual formulation of
non-abelian spin models. Here we give two such formulations which appear 
to be quite different from the conventional formulation mentioned above. 
In our opinion the most essential advantages of our formulations 
are that 1) one of them can be definitely used for the Monte-Carlo 
simulations in $2D$ and 2) they are much more suitable for an analytical 
investigation of the model in the low-temperature region. The second fact 
follows from the properties of the link formulation and   
we refer for the details of why it is so to our papers \cite{su2,2ddual}. 
In the last of those papers we have already 
presented a dual of $2D$ $SU(2)$ spin model and proposed approximate 
representation for the dual partition function at low temperatures. 
Another important feature of one of our formulations is that it carries 
a close analogy with the abelian $U(1)$ model. For instance, 
for $2D$ $SU(2)$ model there are 3 independent degrees of freedom per 
(dual) site. At low temperatures it possesses an analog of the 
vortex--spin-wave representation, etc. 
The deficiency of our approach is that it can be straightforwardly applied 
only to models in which spins are elements of the Lie group. 
We do not know any obvious generalization of link formulation to, say, 
$O(N)$ nonlinear sigma models.

Low-temperature properties of $2D$ non-abelian models are crucial
for construction of their continuum limit. It is commonly recognized
that models possess no phase transition, correlation function has
exponential decay at any coupling and models are asymptotically free.
Despite being more than twenty five years old this expectation has not been
proven rigorously. On the contrary, certain percolation
theory arguments suggest that all
non-abelian models have soft low-temperature phase with power-like decay
of the correlation function \cite{seiler}. It is thus another important
motivation of the present investigation, namely to get deeper
insight into the nature of the mass gap in non-abelian spin models.
For example, in many papers devoted to $2D$ non-abelian models
it is written that ``there is a nonperturbative mass
gap generation at arbitrarily small couplings''. It is not really clear,
however what is precise meaning of this ``nonperturbative generation''. 
It cannot be a simple consequence of 
the link decorrelation which happens in $1D$ models.
Then, one could ask if this ``nonperturbative generation'' follows from
the existence of some non-trivial background of defects like vortices of
the $XY$ model or is due to the strong but smooth disorder of non-abelian
spins, e.g. like center vortices \cite{mgsu2}. As is well known, the dual
formulation of abelian models have been extremely useful in clarifying
all these important physical problems \cite{rigbkt,dualu1,3du1lgt}.
We think that the dual formulation given here is a good starting point
for obtaining reliable analytical results in the low-temperature limit
of non-abelian models.
This is our next goal to develop a technique within dual formulation of
non-abelian models which would allow to investigate them in the limit of
the weak coupling. In this context we study here the following two approaches 
on the simplest example of $SU(2)$ spin model 

\begin{itemize}

\item
We derive an asymptotic expansion of 
the dual Boltzmann weight when $\beta\to\infty$
uniformly valid in fluctuations of the dual variables.
The same procedure is also done for the two-point correlation function. 
As will be seen, at low temperatures the Boltzmann weight converges to a 
certain Gaussian ensemble -- the fact which ensures a possibility to define 
an analog of the vortex--spin-wave representation for the partition and 
correlation functions in the semiclassical limit.
This step provides a major simplification of the model as $\beta\to\infty$
but even in this case it cannot be yet solved exactly.
The formulation obtained permits very simple calculation of the leading 
perturbative contribution to the correlation function. This contribution 
results in the power-like decay of the correlation function. 

\item
The second approach consists in replacing $SU(2)$ matrix elements by 
$ISO(2)$ ones in the vicinity of the identity element of $SU(2)$. 
We shall give a proof that such replacement is valid at low temperatures, 
compute the asymptotics of the dual Boltzmann weight and present simple 
numerical evidence that 
practically for all configurations this way of calculations produces very 
reasonable approximation for the original Boltzmann weight. 

\end{itemize}

This paper is organized as follows. In the section 2 we review briefly the 
conventional dual transformations. After introduction of the link 
formulation we develop original approach to the dual transformations 
and present two forms of the duals of the principal chiral model. 
In the section 3 we investigate the low-temperature properties of 
the dual Boltzmann weights of $SU(2)$ model in two dimensions along 
the lines described above. Some technical details of this investigation 
are given in the appendix. In the section 4 we summarize our results and 
outline some perspectives for future investigations. Also, we calculate 
here the leading perturbative contribution to the correlation function.

\section{Principal chiral models and their duals} 

We work on a $D$-dimensional hypercubic lattice $\Lambda\in Z^d$ 
with lattice spacing $a=1$ and a linear extension $L$. We impose 
either free or periodic boundary conditions (BC). Let $G=U(N), SU(N)$; 
$U(x)\in G$ and $DU(x)$ denote the Haar measure on $G$. $\chi_r(U)$ and $d(r)$ 
will denote the character and dimension of irreducible representation 
$\{ r \}$ of $G$, correspondingly. We treat models with local interactions, 
i.e. interactions between nearest neighbours. Let $H[U]$ be a real invariant 
function on $G$ such that 
\begin{equation}
\mid H[U] \mid  \ \leq H(I)
\label{huleq}
\end{equation}
for all $U$ and coefficients of the character expansion of 
$\exp (\beta H)$ 
\begin{equation}
C[r] \ = \ \int DU \ \exp \left ( \beta H[U] \right ) \chi_r(U)
\label{charexpdef}
\end{equation}
exist. The partition function (PF) of the 
principal chiral model with the symmetry group $G\times G$ is defined as 
\begin{equation}
Z_{\Lambda}(\beta ) = \int \prod_{x\in \Lambda} DU (x)
\exp \left [ \beta \sum_{x\in\Lambda} \sum_{n=1}^D  
H \left [ U(x)U^{\dagger}(x+e_n) \right ]  \right ] \ . 
\label{PFdef}
\end{equation}

\subsection{Standard dual transformations}

Conventional dual transformations can be defined as a sequence
of transformations consisting of the following steps:

\begin{enumerate}

\item
Fourier expansion of the Boltzmann weight $\exp (H)$.
This is an essentially character expansion on the group. 

\item 
Exact integration over original degrees of freedom.
As a result one obtains a set of constraints on the summation
variables (which label representations of the group $G$ and 
matrix elements of $U(x)$) in the character expansion.

\item
Solution of the constraints in terms of new dual variables.

\end{enumerate}
Performing the first step one finds for the PF 
\begin{equation}
Z_{\Lambda}(\beta ) = \prod_{l\in\Lambda}\left [ \sum_{\{ r(l) \}} \ 
\sum_{m(l),k(l)=1}^{d(r)} \ C[r(l)] \right ] \prod_{x\in\Lambda} Q(x) \ , 
\label{PFst1}
\end{equation}
where $l=(x,e_n)$ denotes links of the lattice and $Q(x)$ is a group integral. 
Let $D_r^{mk}[U]$ be a matrix element of $U$ in representation $r$. 
Then $Q(x)$ can be written as  
\begin{equation}
Q(x) \ = \ \int DU \prod_{n=1}^D \left [ D_{r(x,e_n)}^{m(x,e_n)k(x,e_n)}[U] 
D_{r(x-e_n,e_n)}^{m(x-e_n,e_n)k(x-e_n,e_n)}[U^{\dagger}] \right ] \ . 
\label{Qxdef}
\end{equation}
The second step in the most general settings (for arbitrary graph and for 
all $G$) has been accomplished in \cite{pfeiffer}. Through the rest of this 
subsection we consider only the simplest cases of $G=U(1)$ and $G=SU(2)$ 
in two dimensions. These examples already show a marked difference between 
abelian and non-abelian models, on the one hand. On the other hand, we want 
to have relatively simple formulae which can be compared with our formulations. 
For the $U(1)$ model we have 
\begin{equation}
Q(x) \ = \ \delta \left ( \sum_{n=1}^2 
\left [ r(x,e_n)-r(x-e_n,e_n) \right ] \right ) \ ,
\label{Qxu1}
\end{equation}
where $\delta (x)$ is the Kronecker delta. In $SU(N)$ case the result for 
$Q(x)$ in \cite{pfeiffer} is expressed in terms of the Haar intertwiners. We 
re-state this result in more familiar form expanding $Q(x)$ in 
the Clebsch-Gordan (CG) series 
\begin{equation}
Q(x) \ = \ \sum_{J,s,t} \frac{1}{d(J)} \ C_{r_1 m_1 \ r_2 m_2}^{J s} \ 
C_{r_1 k_1 \ r_2 k_2}^{J t} \ C_{r_3 m_3 \ r_4 m_4}^{J t} \ 
C_{r_3 k_3 \ r_4 k_4}^{J s} \ ,
\label{Qxsu2}
\end{equation}
where $r_1=r(x,e_1)$, $r_2=r(x,e_2)$, $r_3=r(x-e_1,e_1)$, $r_4=r(x-e_2,e_2)$ 
and similar notations are used for $m_i$ and $k_i$. 

The third step can be readily accomplished for the $U(1)$ model. We skip 
all the details of the derivation which are well known and can be found, 
for example in \cite{savit, rigbkt}. The result for the PF reads 
\begin{equation}
Z_{\Lambda}^{XY}(\beta ) = 
\sum_{r(x)=-\infty}^{\infty} \  \prod_{l\in \overset{\star}\Lambda} \ 
\Xi_0^{XY}(l) \ ,
\label{dualXY}
\end{equation}
\noindent
where the product runs over all links of the dual lattice 
$\overset{\star}\Lambda$ and the dual Boltzmann weight is given by
\begin{equation}
\Xi_0^{XY}(l) \equiv \Xi_0^{XY}(r(x)-r(x+e_n);\beta ) =
\int_0^{2\pi}\frac{d\phi}{2\pi}\exp \left [ \beta H(\phi ) +
i(r(x)-r(x+e_n)) \right ] \ ,
\label{dualBWXY}
\end{equation}
\noindent
which in the case of the standard choice
$H=\cos\phi$ leads to
\begin{equation}
\Xi_0^{XY}(l) = I_{r(x)-r(x+e_n)}(\beta )
\label{dualXYstd}
\end{equation}
\noindent
with $I_n(x)$ - the modified Bessel function. 

Unfortunately, situation is much more difficult for non-abelian case. 
$Q(x)$ for $SU(2)$ contains not just one but 4 conditions: 2 Kronecker 
delta's which constrain magnetic numbers $m_i$ and $k_i$ 
\begin{equation}
m_1+m_2=k_3+k_4=s \ , \ m_3+m_4=k_1+k_2=t 
\label{magnumbconst}
\end{equation}
and 2 triangular conditions on representations
\begin{equation}
\mid r_1-r_2 \mid \leq J \leq r_1+r_2 \ , \ 
\mid r_3-r_4 \mid \leq J \leq r_3+r_4  \ .
\label{reprconst}
\end{equation}
We are not aware of any attempts in the literature to resolve these 
constraints simultaneously for all lattice sites. Of course, this does not 
cause much problems for the strong coupling expansion where only 
lowest values of $r(l)$ contribute. But situation becomes hopeless 
in the low temperature region where probably all $r(l)$ become relevant. 
Eventually, this reduces to the problem of finding the appropriate summation 
technique when $\beta\to\infty$ which we could not solve. Moreover, 
$Q(x)$ as defined in (\ref{Qxsu2}) is not strictly positive quantity: 
on a number of configurations it is negative. We do not know how to 
perform resummation over the magnetic numbers to get positive Boltzmann weight. 
Most probably, only complete summation over all magnetic numbers will 
produce positive quantity. 
One of the possible ways of how to proceed further is to go over to 
the triangular lattices identifying representations with the length 
of the sides of triangles and magnetic numbers with the orientation of the 
triangles, and we have undertaken some attempts to go along this line. 
Detailed description of this procedure is however beyond the scope of the 
present paper.

\subsection{Link formulation of principal chiral models}

We turn now to a different approach to duality transformations based
on so-called link representation for partition and correlation functions.
That is, the three steps formulated in the beginning of the previous subsection
are replaced by the following ones

\begin{enumerate}

\item
Change of variables in the partition function (\ref{PFdef})
\begin{equation}
V(l)=U(x)U^{\dagger}(x+e_n) \ .
\label{linkfunc}
\end{equation}
\noindent
Integration over original degrees of freedom.
This integration generates a set of constraints on the link matrices 
$V(l)$ known as the Bianchi identities.

\item
Implementation of constraints into the partition function by making use of
the invariant delta-function on the group. This step introduces new degrees
of freedom which can be associated with plaquettes of the original lattice
and which label the irreducible representations of the group $G$.
At this step one can go over to the dual lattice so that the new variables
belong to the elements of the dual lattice. Integration over link 
variables. These two steps give already some dual version of 
the principal chiral model.

\item
Before integration over link variables one can decouple summations over 
matrix indices employing certain orthogonality relations for the 
CG coefficients (or for some other objects) of the group and multiplying 
the group matrices entering Bianchi identity in some specially ordered way. 
Though details depend on the dimension of the lattice the result is that 
only independent dual variables remain in the theory and their interaction 
is confined to a subset of all hypercubes of the dual lattice.

\end{enumerate}

The first step of this program has been accomplished twenty years
ago in \cite{linkrepr}. We therefore restrict ourselves only to the
description of the result. The PF (\ref{PFdef}) can be exactly
reformulated in terms of link variables as (with periodic BC)
\begin{equation}
Z_{\Lambda}(\beta ) = \int \prod_{l\in\Lambda} dV(l)
\exp \left[ \beta \sum_{l\in\Lambda} H[V(l)]  \right] J[V] \ ,
\label{lPF}
\end{equation}
\noindent
with Jacobian $J[V]$ given by 
\begin{equation}
J[V] =  \prod_{p\in\Lambda} J[V(p)]  \ \prod_{n=1}^D J[H(n)] \ ,
\label{jacobdef}
\end{equation}
where $J[V(p)]$ and $J[H(n)]$ are given by 
\begin{equation}
J[V(p)]  \ = \  \sum_{\{ r \}} d(r) 
\chi_r \left( \prod_{l\in p}V(l) \right)  \ ,
\label{jacob}
\end{equation}
\noindent
\begin{equation}
J[H(n)] \ = \ 
\sum_{\{ r \}} d(r) \chi_r \left( \prod_{l_n=1}^LV(l_n) \right) \ , \ 
l=(x,e_n) \ .
\label{holonconstr}
\end{equation}
\noindent
$\prod_p$ is a product over all plaquettes of lattice $\Lambda$,
the sum over $\{ r \}$ is sum over all representations of $G=U(N),SU(N)$, 
$d(r)=\chi_r(I)$ is the dimension of $r$-th representation. 
The group character $\chi_r$ depends on a product of the link 
matrices $V(l)=V_n(x)$ along a closed path (plaquette in our case):
\begin{equation}
\prod_{l\in p}V(l) = V_n(x)V_m(x+e_n)V_n^{\dagger}(x+e_m)V_m^{\dagger}(x) \ .
\label{prod}
\end{equation}
\noindent
The expression $\sum_r d_r \chi _r(\prod_{l\in p}V(l))$ is the invariant
group delta-function which reflects the fact that the product of
$U(x)U^{\dagger}(x+e_n)$ around plaquette equals $I$
\begin{equation}
\prod_{l\in p}V(l) = I \ .
\label{Bloc}
\end{equation}
\noindent
This constraint is called local Bianchi identity while the constraint on 
holonomy operators 
\begin{equation}
\prod_{l_n=1}^LV(l_n) = I \ 
\label{Bglob}
\end{equation}
\noindent
is usually called global Bianchi identity. The only solution of two constraints 
(\ref{Bloc}) and (\ref{Bglob}) is a pure gauge (\ref{linkfunc}) which 
restores the equivalence of the standard and the link formulations. 
On the lattice with free BC the global constraint should be omitted. 
Furthermore, in the thermodynamic limit (TL) the contribution from 
the constraints on the holonomy operators vanishes very fast. 
In the perturbation theory in $2D$ it vanishes like inverse powers of $L$ 
\cite{su2}, and similar behaviour is expected for $D>2$. Since we are 
eventually interested in the TL we neglect these global constraints 
(in fact, the presence of global constraints does not cause any principal 
problems as will be seen from the procedure described below but their absence 
makes the technical details somewhat simpler). Thus, for both free and 
periodic BC we work with the PF
\begin{equation}
Z_{\Lambda}(\beta ) = \int \prod_{l\in\Lambda} dV(l)
\exp \left[ \beta \sum_{l\in\Lambda} H[V(l)]  \right] 
\prod_{p\in\Lambda} J[V(p)]  \ .
\label{linkPFfin}
\end{equation}
\noindent
The two-point correlation function in representation $j$ takes the form 
\begin{equation}
\Gamma_j (x,y) = \left \langle d^{-1}(j)
\chi_j \left ( \prod_{l\in C_{xy}}W(l) \right ) \right \rangle \ ,
\label{corfuncdef}
\end{equation}
where $C_{xy}$ is some path connecting points $x$ and $y$ and 
$W(l)=V(l)$ if along the path the link $l$ points positive direction 
and $W(l)=V^{\dagger}(l)$, otherwise. 
For more details on the link formulation we refer the reader to our paper
\cite{su2}, where we have developed a weak coupling expansion for $SU(N)$
spin models using the link representation.

\subsection{Dual of the link formulation: I}

Here we are going to accomplish the step 2 described in the previous 
subsection. The idea that the true dual variables in the link formulation 
are the Fourier conjugate to the local Bianchi identity has been proposed 
also in \cite{linkrepr} but the formulae have been explicitely written 
only for abelian $U(1)$ model in $2D$. We thus want to reformulate the model 
(\ref{linkPFfin}) on the dual lattice only in terms of discrete variables 
that are in our case representations $\{ r(p) \}$ and magnetic quantum numbers 
$m_i(p)$. We shall use the following conventions for the dual lattice. 
$\Omega_x$ will denote an object dual to site $x$, i.e. $D$-cell of the 
dual lattice, $\Omega_l$ - $(D-1)$-cell of the dual lattice, 
$\Omega_p$ - $(D-2)$-cell of the dual lattice and so on. By definition, 
object dual to k-th cell is $(D-k)$-cell. 
As follows from (\ref{linkPFfin}) and from the definition of the 
character which enters the group delta-function in (\ref{jacob}) 
$\chi_r(V) = \sum_n V_r^{nn}$
on the dual lattice the PF may be written as
\begin{equation}
Z_{\Lambda}(\beta ) = 
Z_{\overset{\star}\Lambda}(\beta ) = \sum_{\{ r(\Omega_p) \}}\prod_{\Omega_p}
\left [ d[r(\Omega_p)] \sum_{m_i(\Omega_p)}^{d[r(\Omega_p)]} \right ] 
\prod_{\Omega_l} \Xi_0(\Omega_l) \ .
\label{PFsundual}
\end{equation}
\noindent
Due to the trace there are 4 variables $m_i$ at each $\Omega_p$,
thus $i=1,2,3,4$. The dual weight $\Xi_0(\Omega_l)$ is given by the following
one-link integral
\begin{equation}
\Xi_0(\Omega_l) \ = \ 
\int dV e^{\beta H[V] } \ \prod_{\nu =1}^{D-1} \left [ 
V_{r(\Omega_p)}^{m_im_{i+1}} \ 
V_{r(\Omega_p^{\prime})}^{\dagger \ n_in_{i+1}} \right ]_{\nu}  \ ,
\label{dualweight}
\end{equation}
\noindent
where $V_{r}^{mn}$ is a matrix element of $r$-th representation. 
$(D-2)$-cells $\Omega_p$ and $\Omega_p^{\prime}$ share $(D-1)$-cell 
$\Omega_l$. $\prod_{\nu}$ runs over $(D-1)$ such pairs. 

Similar form for the two-point correlation function in
the representation $j$ reads
\begin{eqnarray}
\Gamma_j (x,y) =  \frac{Z_{\Lambda}^{-1}(\beta )}{d(j)} \ 
\sum_{s_1}^{d(j)} ...  \sum_{s_R}^{d(j)} \ \sum_{\{ r(\Omega_p) \}}\prod_{\Omega_p}
\left [ d[r(\Omega_p)] \sum_{m_i(\Omega_p)}^{d[r(\Omega_p)]} \right ] 
\nonumber  \\ 
 \times \prod_{\Omega_l\in \overset{\star}C_{xy}} \Xi^{s_is_{i+1}}_j(\Omega_l) \ 
\prod_{\Omega_l\notin \overset{\star}C_{xy}} \Xi_0(\Omega_l) \ ,
\label{2func}
\end{eqnarray}
\noindent
where $s_{R+1}=s_1$ and the link integral on 
$\Omega_l\in \overset{\star}C_{xy}$ is
\begin{equation}
\Xi^{s_is_{i+1}}_j(\Omega_l) \ = \ \int dV e^{\beta H[V] } \ 
\prod_{\nu =1}^{D-1} \left [ V_{r(\Omega_p)}^{m_im_{i+1}} \ 
V_{r(\Omega_p^{\prime})}^{\dagger \ n_in_{i+1}} \right ]_{\nu} \ V_j^{s_is_{i+1}} \ .
\label{2funcdualw}
\end{equation}
\noindent
Here, $\overset{\star}C_{xy}$ is a path dual to the path $C_{xy}$ between 
points $x$, $y$ on the original lattice. 

To vizualize last formulae let us give expression for the PF 
in two dimensions with $G=SU(2)$. In $2D$ we have $\Omega_x=p$, 
$\Omega_l=l$ and $\Omega_p=x$. Placing as usually dual sites in the centers 
of original plaquettes we obtain 
\begin{equation}
Z_{\overset{\star}\Lambda}^{SU(2)}(\beta ) \ = \ 
\sum_{r(x)=0,\frac{1}{2},1,... }^{\infty}
\ \prod_{x\in\overset{\star}\Lambda}
\left [ (2r(x) + 1) \sum_{m_i(x)=-r(x)}^{r(x)} \right ] 
\prod_{l\in\overset{\star}\Lambda} \Xi_0(l) \ ,
\label{PFLsu2}
\end{equation}
\noindent
where the dual weight 
\begin{equation}
\Xi_0(l) \ \equiv \ 
\Xi_0 \left ( r(x),m_i(x),m_{i+1}(x);r(x+e_n),t_i(x+e_n),t_{i+1}(x+e_n); 
\beta \right ) 
\label{dwdefsu2}
\end{equation}
\noindent
becomes 
\begin{equation}
\Xi_0(l) \ = \ 
\int dV e^{\beta  H[V] } \ V_{r(x)}^{m_i(x) \ m_{i+1}(x)} \ 
V_{r(x+e_n)}^{\dagger \ t_i(x+e_n) \ t_{i+1}(x+e_n) } \ .
\label{linkint}
\end{equation}
\noindent

It is very easy to make an integration in (\ref{linkint}) expanding the result
into the CG series. One then finds the following representation
for the dual weight of $SU(2)$ model
\begin{equation}
\Xi_0 (l) =
\frac{1}{2r_2+1} \sum_{J,k} C[J] \ C_{r_1 m_1 \ J k}^{r_2 t_2} \ 
C_{r_1 m_2 \ J k}^{r_2 t_1} \ ,
\label{xio2dexact}
\end{equation}
\noindent
where we have denoted $r_1=r(x)$, $r_2=r(x+e_n)$, $m_1=m_i(x)$ and so on. 
For the standard choice $H[V]=\chi_{1/2}(V)$ the coefficients $C[J]$ can 
be computed explicitely 
\begin{equation}
C[J] \ = \ \frac{2J+1}{\beta} I_{2J+1}(2\beta) \ .
\label{CJ2d}
\end{equation}
\noindent
For the correlation function in Eq.(\ref{2func}) using the CG
expansion one gets
\begin{equation}
\Xi_j^{s_i s_{i+1}}(l) =
\sum_{J \alpha_1 \alpha_2} C_{r_1 m_1 \ j s_i}^{J \alpha_1} \ 
C_{r_1 m_2 \ j s_{i+1}}^{J \alpha_2} \ 
\Xi_0 (J, \alpha_1, \alpha_2; r_2 ,t_1,t_2; \beta ) \ .
\label{xij2d}
\end{equation}
\noindent
Last expressions can be compared with the dual of the $U(1)$ model given 
in (\ref{dualXY}) and with the standard dual formulation of $SU(2)$ model 
described by formulae (\ref{PFst1}) and (\ref{Qxsu2}). Certainly, the dual 
weight for $SU(2)$ model $\Xi_0(l)$ as given in (\ref{linkint})-(\ref{CJ2d}) 
is very close to what we have for the dual weight in $U(1)$ model: 1) dual 
variables, i.e. representations $r(x)$ and magnetic numbers $m_i(x)$, live 
on sites of the dual lattice; 2) the full Gibbs measure is factorized into 
product of dual weights $\Xi_0(l)$ over links of the dual lattice. 
Unlike the standard dual formulation there is no complicated triangular 
conditions on representations: triangular constraint on $J$ in 
(\ref{xio2dexact}) is only one-link problem which does not lead to any 
significant complications. Nevertheless, there is also a difference from 
the abelian case. As follows from the properties of the CG coefficients we 
have one constraint on the magnetic numbers on every link, namely summation 
over $k$ in (\ref{xio2dexact}) produces the following condition
\begin{equation}
m_i(x)-m_{i+1}(x) \ = \ t_{i+1}(x+e_n)-t_i(x+e_n) \ .
\label{mgnconstsu2}
\end{equation}
This shows that not all dual variables are in fact independent, and this 
is the main difference from the $U(1)$ model. There is no simple way to 
make a change of variables such that one gets only independent variables and 
keeps the locality of interaction. In the next subsection we present 
a different way of the construction of the dual theory from the link formulation 
in such a way that the dual weight depends only on independent dual variables. 

We finish this subsection with the brief description of some important 
features of $\Xi_0(l)$:

\begin{itemize}

\item
As follows from the properties of the coefficients of the expansion
$C[J]$ the series in $J$ in (\ref{xio2dexact}) gives directly
the strong coupling expansion of the model written in closed and compact form. 
Much less trivial task is to get weak coupling expansion for $\Xi_0(l)$
since all $J$ in the series (\ref{xio2dexact}) become relevant.

\item
On all configurations $\{ r_i,m_i,n_i \}$ $\Xi_0 (l)$
is strictly positive $\Xi_0 (l) > 0$. Though we could not prove it
rigorously this claim is supported by the following facts:
1) the first term in the strong coupling expansion is strictly positive, thus
at sufficiently small $\beta$ where the series converges very fast
$\Xi_0 (l)$ is positive; 2) the leading term of the asymptotic expansion
of $\Xi_0 (l)$ at large $\beta$ is strictly positive on all configurations;
3) numerical computations of $\Xi_0 (l)$ on a number of configurations
and in a wide region of $\beta$ also support this conclusion (we have checked 
this statement using Mathematica on more than $10^4$ configurations).
If $\Xi_0 (l) > 0$ on all configurations this gives a chance for a numerical
Monte-Carlo simulations of the dual model.

\item
The dominant contribution to $\Xi_0 (l)$ at large $\beta$ comes from diagonal
components of the rotation matrices, non-diagonal contribution is suppressed
roughly as $[(m-n)!]^{-1}$. This is, of course a consequence of the fact
that when $\beta\to\infty$ the link matrix performs only small fluctuations
around unity. In turn, this property gives a possibility to compute
low-temperature asymptotic expansion of $\Xi_0 (l)$. This will be 
subject of the next section.

\end{itemize}

\subsection{Dual of the link formulation: II}

The main idea of the following approach is to introduce the independent 
dual variables from the beginning, namely to find a representation 
for the group invariant delta-function in terms of quantities which would 
be free of any conditions. We explain the realization of this idea for 
$2D$ $SU(2)$ model and then show how it can be extended to other groups 
and to higher dimensions. 

We start with the representation for $SU(2)$ matrix
elements in which the dependence on the magnetic numbers enters only
through the Clebsch-Gordan coefficients $C^{a\alpha}_{b\beta \ c\gamma}$
\footnote{In what follows 
we use notations and conventions of Ref.\cite{matrelem} for the
Clebsch-Gordan coefficients, spherical harmonics, etc.}
\begin{equation}
V_r^{mn}(\omega , \theta , \phi ) = \sum_{\lambda k}
C^{rn}_{rm\ \lambda k} \ U_r^{\lambda k}(\omega , \theta , \phi ) \ ,
\label{Vrmn}
\end{equation}
where $U_r^{\lambda k}(\omega , \theta , \phi )$ is expressed through
the spherical harmonics ${\rm Y}_{\lambda k} (\theta , \phi)$
and generalized characters $\chi^r_{\lambda}(\omega )$ of $SU(2)$
\begin{equation}
U_r^{\lambda k}(\omega , \theta , \phi ) = (-i)^{\lambda}
\frac{2\lambda +1}{2r+1} \sqrt{\frac{4\pi}{2\lambda +1}} \
{\rm Y}_{\lambda k} (\theta , \phi) \ \chi^r_{\lambda}(\omega ) \ .
\label{su2matr}
\end{equation}
Generalized character of rank $\lambda$ in representation $r$
can be defined through the relation
\begin{equation}
\chi^r_{\lambda}(\omega ) = i^{\lambda}\sum_{m=-r}^{r}e^{-im\omega }
C^{rm}_{rm\ \lambda 0} \ .
\label{genchar}
\end{equation}
We remind that the invariant measure in this parameterization is
\begin{equation}
\int dV = \frac{1}{4\pi^2}\int_0^{2\pi}\sin^2\frac{\omega}{2}d\omega
\int_0^{\pi}\sin\theta d\theta \int_0^{2\pi}d\phi \
\label{invmeassu2}
\end{equation}
and the fundamental trace becomes
\begin{equation}
{\rm Tr}V_{1/2} = 2\cos\frac{\omega}{2} \ .
\label{trv}
\end{equation}
The basic formula which we need is the following representation for $SU(2)$ 
characters 
\begin{equation} 
\chi_r \left ( V_1V_2 \right ) \ = \ \frac{(2 r+1)}{4\pi} 
\int_0^{2\pi}d\varphi \int_0^\pi d\alpha \sin\alpha \ 
\Phi (r,\varphi , \alpha ; V_1) \ \Phi (r, \varphi , \alpha ; V_2) \ ,
\label{su2char}
\end{equation}
\begin{equation} 
\Phi (r,\varphi , \alpha ; V ) \ = \ \sum_{\lambda =0}^{2r} (-i)^\lambda \ 
\frac{2\lambda + 1}{2 r + 1} \ P_{\lambda}(t) \ 
\chi_{\lambda}^r \left ( \omega \right ) \ ,
\label{Phidef}
\end{equation} 
where $P_{\lambda}(t)$ is the Legendre polynomial and 
\begin{equation}
t = \cos\theta \cos\alpha + 
\sin\theta\sin\alpha\cos(\phi - \varphi) \ .
\label{combangle}
\end{equation}
This representation can be easily proved if one uses orthogonality relations 
for the spherical harmonics and then the addition theorem for the generalized 
characters. 

As the next step, we divide all sites of the dual lattice into sets of even 
and odd sites. Site $x=(x_1,x_2)$ is even if both $x_1$ and $x_2$ are even or 
odd, and is odd if one of $x_i$ is odd. Thus, all even sites are surrounded by 
odd sites and vice-versa. Using cyclic properties of the trace we can couple the 
four matrices entering local Bianchi identities in two different ways, namely 
\begin{equation}
\chi_r(V_x) \ = \ \chi_r\left [ V_{12}V_{34} \right ]  \ , 
\label{chareven}
\end{equation}
if the site $x$ is even and 
\begin{equation}
\chi_r(V_x) \ = \ \chi_r\left [ V_{23}V_{41} \right ]  \ , 
\label{charodd}
\end{equation}
if the site $x$ is odd. We introduced here obvious notations 
\begin{equation}
V_{12} = V(l_1)V(l_2), \ V_{34} = V^{\dagger}(l_3)V^{\dagger}(l_4), \ 
V_{23} = V(l_2)V^{\dagger}(l_3), \ 
V_{41} = V^{\dagger}(l_4)V(l_1). 
\label{notVinsite}
\end{equation}

\begin{figure}[ht]
\centerline{\epsfxsize=6cm \epsfbox{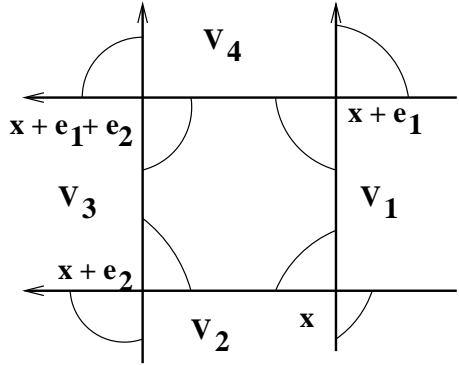}}
\caption{\label{matrcoupl} 
Dual even plaquette and coupling of matrices in group delta's. $V_i\equiv V(l_i)$, 
sites $x$ and $x+e_1+e_2$ are even, while $x+e_1$ and $x+e_2$ are odd.}
\end{figure}

With help of (\ref{su2char}) and (\ref{chareven}), (\ref{charodd}) the Jacobian 
$J[V]$ defined in (\ref{jacobdef}) and (\ref{jacob}) can be written on 
the dual lattice as 
\begin{equation}
J[V] =  \prod_{x, \mbox{even}} J[V(x)]  \ 
\prod_{x, \mbox{odd}} J[V(x)] \ ,
\label{jacobdual}
\end{equation}
where in even sites 
\begin{equation} 
J(V_x) \ = \ \sum_{r = 0,1/2,..}\frac{(2 r+1)^2}{4\pi} 
\int_0^{2\pi}d\varphi \int_0^\pi d\alpha \sin\alpha \ 
\Phi (r,\varphi , \alpha ; V_{12}) \ \Phi (r, \varphi , \alpha ; V_{34}) \ ,
\label{su2deltaeven}
\end{equation}
while for odd sites 
\begin{equation} 
J(V_x) \ = \ \sum_{r = 0,1/2,..}\frac{(2 r+1)^2}{4\pi} 
\int_0^{2\pi}d\varphi \int_0^\pi d\alpha \sin\alpha  \ 
\Phi (r,\varphi , \alpha ; V_{23}) \ \Phi (r, \varphi , \alpha ; V_{41}) \ .
\label{su2deltaodd}
\end{equation}
A set of variables $\xi (x) = \lbrace r(x), \alpha (x), \varphi (x)\rbrace$ 
is chosen to be dual variables at site $x$. Finally, we decompose all 
plaquettes of the dual lattice into sets of even and odd plaquettes in such 
a way that all even plaquettes are surrounded by odd ones and vice-versa. 
Due to the special coupling of matrices as described 
above, the integration over link matrices is now coupled only within every, 
for example even plaquette. This is shown in Fig.\ref{matrcoupl}. 

Substituting (\ref{su2deltaeven}) and  (\ref{su2deltaodd}) in (\ref{lPF}), 
re-denoting notations for links $l\in p$ uniformly for all even plaquettes 
as shown in Fig.\ref{matrcoupl} we obtain 
dual expression for the partition function 
\begin{eqnarray}
Z_{\overset{\star}\Lambda}^{SU(2)}(\beta ) \ &=& \ 
\prod_{x\in\overset{\star}\Lambda} \left [ 
\sum_{r(x) = 0,1/2,...}^{\infty} 
(2 r(x) + 1)^2 \int_0^{2\pi} \frac {d \varphi (x)}{4 \pi} 
\int_0^{\pi} d\alpha (x) \sin\alpha (x) \right ] \ \nonumber  \\ 
&\times& \prod_{p,\mbox{even}} B_0\left [ \xi (x),\xi (x+e_1), 
\xi (x+e_2), \xi (x+e_1+e_2); \  \beta \right ]   \ ,
\label{dual2su2}
\end{eqnarray}
where the dual Boltzmann weight 
\begin{equation}
\label{dualBWdef}
B_0(p) \ \equiv \ B_0\left [ \xi (x),\xi (x+e_1),\xi (x+e_2),\xi (x+e_1+e_2); \  
\beta \right ]  
\end{equation}
is given by 
\begin{eqnarray}
\label{dualBW}
B_0(p) \ &=& \ \int \prod_{i=1}^4 dV_i 
\exp \left \{ \beta\sum_{i=1}^4H[V_i] \right \} \ \Phi (\xi (x);V_1 V_2) \\ 
&\times& \Phi (\xi (x+e_2);V_2^{\dagger}V_3) \ 
\Phi (\xi (x+e_1+e_2);V_3^{\dagger} V_4^{\dagger}) \ 
\Phi (\xi (x+e_1);V_4 V_1^{\dagger}) \ .  \nonumber 
\end{eqnarray}
Dual expression for the correlation function can be easily re-constructed 
from Eq.(\ref{corfuncdef}) and last formulae. Consider, for simplicity the 
shortest path between points $x=(0,0)$ and $y=(0,R)$. The corresponding dual 
path is a set of links which point direction $e_1$. Let $\mid x-y \mid = R$ be 
even. Since every link belongs to one and only one even plaquette we obtain 
\begin{eqnarray}
\Gamma_j (x,y) &=&  \frac{Z_{\Lambda}^{-1}(\beta )}{2j+1} \ 
\prod_{x\in\overset{\star}\Lambda} \left [ 
\sum_{r(x) = 0,1/2,...}^{\infty} 
(2 r(x) + 1)^2 \int_0^{2\pi} \frac {d \varphi (x)}{4 \pi} 
\int_0^{\pi} d\alpha (x) \sin\alpha (x) \right ] \ \nonumber  \\ 
&\times& \prod_{p^{\prime}} B_0(p^{\prime}) \ 
\sum_{s_1=-j}^{j} \cdots  \sum_{s_R=-j}^{j} \ 
\prod_{p(C)} B_j(p(C);s_{i-1}s_is_{i+1})   \ ,
\label{corrfd1}
\end{eqnarray}
\noindent
where $p^{\prime}$ denotes a set of even plaquettes which do not contain 
links $l\in\overset{\star}C_{xy}$ and $p(C)$ - a set of even plaquettes 
which contain such links. In notations of Fig.\ref{matrcoupl} we have 
\begin{eqnarray}
\label{dualBW1}
&&B_j(p(C);s_{i-1}s_is_{i+1}) \ = \ \int \prod_{i=1}^4 dV_i 
\exp \left \{ \beta\sum_{i=1}^4H[V_i] \right \} \ V_j^{s_{i-1}s_i}(l_1)  \ 
V_j^{s_{i}s_{i+1}}(l_3) \\
&& \times \Phi (\xi (x);V_1 V_2) \ 
\Phi (\xi (x+e_2);V_2^{\dagger}V_3) \ 
\Phi (\xi (x+e_1+e_2);V_3^{\dagger} V_4^{\dagger}) \ 
\Phi (\xi (x+e_1);V_4 V_1^{\dagger}) \ .  \nonumber 
\end{eqnarray}

Finally, the disorder operator in the dual representation can be readily 
obtained from its expression in the link formulation. Let 
$H[-V(l)]=-H[V(l)]$. Consider the following disorder operator  
\begin{equation}
D(x,y) \ = \ 
\exp \left [  2\beta\sum_{l\in T} H \left [ -V(l) \right ] \right ] \ , 
\label{disparamdef}
\end{equation}
where 
$T=\{ l \} =  [(x,e_1),(x+e_2,e_1),...,(y-e_2,e_1),(y,e_1)]$. Expectation value 
of this operator can be written as a ratio of two partition functions. 
With help of the change of variables $V(l)\to -V(l)$ this ratio can be 
presented as the following expectation value in the dual formulation
\begin{equation}
\langle D(x,y) \rangle \ = \ 
\left \langle (-1)^{2r(x)-2r(y)} \right \rangle  \ . 
\label{disparamdual}
\end{equation}

Generalization of these representations to higher dimensions is straightforward. 
In this case we couple link matrices in such a way as to decouple integration 
over them into a set of integrations within each even $D$-cell of the dual 
lattice. Then the partition function can be written as 
\begin{eqnarray}
Z_{\overset{\star}\Lambda}^{SU(2)}(\beta ) \ &=& \
\prod_{\Omega_p}\left [ \sum_{r(\Omega_p)=0,1/2,...}^{\infty}  
(2 r(\Omega_p) + 1)^2 \int_0^{2 \pi} \frac {d \varphi (\Omega_p)}{4 \pi} 
\int_0^{\pi} \sin \alpha (\Omega_p) d \alpha(\Omega_p) \right ] \nonumber  \\ 
&\times& \prod_{h_D, \mbox{even}} B(h_D) \ ,
\label{dualsu2D}
\end{eqnarray}
where the dual Boltzmann weight reads 
\begin{equation}
B(h_D) = \int \prod_{\Omega_l \in h_D} 
\left[ d V(\Omega_l) 
\exp \left \{ \beta H [V(\Omega_l)] \right \} \right] 
\prod_{\Omega_p \in h_D} \Phi (\xi (\Omega_p); \gamma (\Omega_p)) \ .
\label{dualBWDef}
\end{equation}
Here, $\xi (\Omega_p) = \lbrace r(\Omega_p), \alpha(\Omega_p), 
\varphi(\Omega_p) \rbrace$ are dual variables which belong to $(D-2)$-cells 
of the dual lattice and  $\gamma (\Omega_p) = \lbrace \omega (\Omega_p), 
\theta(\Omega_p), \Phi(\Omega_p) \rbrace$ are combined angles arising from 
multiplication of $(D-1)$-cell matrices (original link matrices ) which 
belong to the same plaquette trace. 

To our opinion, the dual representation presented above bears the closest 
analogy to the abelian case. For example, in two-dimensional model there are three 
independent dual variables per site as could be expected on general grounds. 
Further, as follows from the Eq.(\ref{dualBW}) the interaction between dual 
variables is local though it is not exactly the same as in abelian models: 
in addition to nearest-neighbour interaction, there is an interaction over 
diagonals of even plaquettes. The deficiency of this dual representation is 
that the dual weight is in general complex valued. This makes the numerical 
MC computations impossible. On the other hand, we believe it is well suited 
for an analytical investigation of the model at low temperatures. 

In the described above scheme the realization of ideas leading to the local 
dual formulation relies on the special representation for $SU(2)$ matrices 
(\ref{Vrmn}), (\ref{su2matr}). This representation appears to be very efficient 
at large $\beta$ as we shall show in the next section. We do not know if similar 
parameterization exists for any group $G$. Therefore, for arbitrary group $G$ we 
proceed as follows. Instead of (\ref{su2char}) we use unitarity relations 
for the CG coefficients of $U(N)$ or $SU(N)$ group \cite{klymik} 
\begin{equation}
\sum_{\{ Jk \}}  C^{Jk}_{r_1m_1 \ r_2m_2} \ 
C^{Jk}_{r_1n_1 \ r_2n_2} \ = \ \delta_{m_1,n_1} \delta_{m_2,n_2} 
\label{unitCG}
\end{equation}
which allows to present group characters as 
\begin{equation} 
\chi_r \left ( V_1V_2 \right ) \ = \ \sum_{\{ Jk \}} 
G(r,J,k;V_1) \ G(r,J,k;V_2) \ ,
\label{suNchar}
\end{equation}
where
\begin{equation} 
G(r,J,k;V) \ = \ \sum_{m_1,m_2} C^{Jk}_{rm_1 \ rm_2} \ V_r^{m_1m_2}\ .
\label{suNGdef}
\end{equation}
We then proceed exactly as for $SU(2)$. For the sake of simplicity let us 
consider the two-dimensional theory. Identifying 
$\xi (x) = \{ r(x), J(x), k(x) \}$ as new dual variables 
the PF can be written as 
\begin{equation}
Z_{\overset{\star}\Lambda}^{SU(N)}(\beta ) \ = \
\sum_{\xi (x)} \prod_{x\in\overset{\star}\Lambda} d[r(x)] 
\prod_{p,\mbox{even}} \ B(p) \ , 
\label{dualsuN}
\end{equation}
where the dual Boltzmann weight takes the form 
\begin{eqnarray}
\label{dualsuNBW}
B(p) \ &=& \ \int \prod_{i=1}^4 dV_i 
\exp \left \{ \beta\sum_{i=1}^4H[V_i] \right \} \ G (\xi (x);V_1 V_2) \\ 
&\times& G (\xi (x+e_2);V_2^{\dagger}V_3) \ 
G (\xi (x+e_1+e_2);V_3^{\dagger} V_4^{\dagger}) \ 
G (\xi (x+e_1);V_4 V_1^{\dagger}) \ . \nonumber
\end{eqnarray}
Substituting (\ref{suNGdef}) into the last equation one can express $B(p)$ 
in terms of one-link integrals defined in (\ref{dualweight}). Since 
the CG coefficients of $SU(N)$ group can be chosen real \cite{klymik}, 
the dual weight in this representation is also real. 
We have not studied if it is positive.

\section{$2D$ $SU(2)$: Dual weights at low temperatures}

One of the most important application of the dual formulation is 
the investigation of the low-temperature region of two-dimensional 
non-abelian models. The first step in this investigation is to establish 
an asymptotic expansion for dual weights at large $\beta$. 
Clearly, it is necessary to get asymptotics uniformly valid in all 
fluctuations of dual variables. 
It turns out that such asymptotics can indeed be constructed, and this is, 
in our opinion one of the most important advantages of our dual formulation.
Calculation of the asymptotic expansion essentially relies on the fact
that when $\beta\to\infty$ the link matrix performs only small fluctuations
around unity both in the finite volume and, most importantly in the
thermodynamic limit. Note, that this is not the case for the original 
($U(x)$) degrees of freedom: in the large volume limit their fluctuations 
are not bounded. 

\subsection{Effective theory at low temperatures}

In this subsection we construct an effective low-temperature theory for 
the dual formulation (\ref{dual2su2}). The corresponding dual weight is 
given by Eq.(\ref{dualBW}). We modify it by introducing sources $j_k(l)$ 
which can be related to the correlation function
\begin{eqnarray}
\label{dualBWsrc}
B_j(p) \ &=& \ \int \prod_{l=1}^4 dV_l 
\exp \left \{ \sum_{l=1}^4 \left [ \beta H[V_l] 
+ j_k(l)\omega_k(l) \right ] \right \} \ 
\Phi (\xi (x);V_1 V_2) \\ 
&\times& \Phi (\xi (x+e_2);V_2^{\dagger}V_3) \ 
\Phi (\xi (x+e_1+e_2);V_3^{\dagger} V_4^{\dagger}) \ 
\Phi (\xi (x+e_1);V_4 V_1^{\dagger}) \ ,  \nonumber 
\end{eqnarray}
where 
$$ 
\omega_1(l)=\sin\frac{\omega}{2}\sin\theta\cos\Phi \ , \ 
\omega_2(l)=\sin\frac{\omega}{2}\sin\theta\sin\Phi \ , \
\omega_3(l)=\sin\frac{\omega}{2}\cos\theta \ .
$$
It is well known (see, e.g. \cite{contest}) 
that when $\beta\to\infty$ the link matrix performs only 
small fluctuations around unit matrix. Our next calculations 
relay on this fact. 
Let $H[V]$ has a unique maximum at $V=I$. We also require that $H[V]$ 
possesses character expansion. Then, in the vicinity $V=I$ and in the 
parameterization  (\ref{Vrmn}), (\ref{su2matr}) one can always construct 
the following expansion 
\begin{equation}
H[V_l] \ = \ H[I] -\frac{\gamma}{2}\sin^2\frac{\omega (l)}{2} 
+ {\cal O}(\omega^3) \ . 
\label{Hexp}
\end{equation}
For the standard choice $H[V]={\rm Tr} V_{1/2}=2\cos\frac{\omega}{2}$ 
one finds $H[I]=2$ and $\gamma = 2$. 
Next, we need uniform asymptotic expansion of the function 
$\Phi (\xi ;V)$. We derive such an expansion in the appendix A. The leading 
term of this expansion takes the form
\begin{equation} 
\ln \Phi (\xi ;V) \ = \ -i\sin\frac{\omega}{2}t\sqrt{R^2-1} + 
{\cal{O}}(\sin^2\frac{\omega}{2}) \ ,
\label{Phiexplead}
\end{equation} 
where the classical angular momentum $R$ is introduced in (\ref{clangm}).
As follows from the semiclassical expansion (\ref{Phisemiclexp}), the 
remainder is actually bounded as ${\cal O}(\beta^{-1})$ for all $R$. 
Therefore, corrections ${\cal{O}}(\sin^2\frac{\omega}{2})$ and higher 
are in fact negligible in the limit $\beta\to\infty$ so we treat them  
perturbatively. Lastly, we substitute all these expressions into 
(\ref{dualBWsrc}), express combined angles $\omega$, $\theta$, $\Phi$ 
in terms of link angles and perform integration. We omit all these simple 
but rather cumbersome calculations and give only the final result.
Introducing notations
\begin{equation}
C(\beta ) =  
\exp \left [ 2\beta L^2H[I] - 3L^2\ln (\gamma\beta ) \right ] \ ,
\label{constant}
\end{equation}
\noindent
\begin{eqnarray}
\label{Rkx}
R_k(x) \ &=& \ 
\frac{1}{\sqrt{\gamma\beta}} \ t_k(x)\sqrt{R^2(x)-1} \ , \\  
t_1(x)&=&\sin\alpha (x)\cos\varphi (x) \ , \ 
t_2(x)=\sin\alpha (x)\ \sin\varphi (x) \ , \ 
t_1(x)=\cos\alpha (x) \  \nonumber  
\end{eqnarray}
and making the change of variables $R(x)=2r(x)+1$ so that $R(x)$ takes integer 
values, we present the result for the dual partition function 
at large $\beta$ in the form 
\begin{equation}
Z_{\overset{\star}\Lambda}(j,\beta ) =  C(\beta ) 
\prod_{x\in\overset{\star}\Lambda} \left [ 
\sum_{R(x) = 1}^{\infty} R^2(x) \int_0^{2\pi} \frac {d \varphi (x)}{2 \pi} 
\int_0^{\pi} d\alpha (x) \sin\alpha (x) \right ]  \exp [S_{eff}] \ .
\label{dualeff}
\end{equation}
An expression for the effective action $S_{eff}$ is too large to be given 
here in full, therefore we present only its expansion.
Rescaling sources as $\tilde{j}_k(l)=\frac{1}{\sqrt{\gamma\beta}}j_k(l)$ 
the effective action can be expanded as 
\begin{equation}
S_{eff} \ = \ S_0 + \frac{i}{\sqrt{\gamma\beta}} \ S_1 + 
\frac{1}{\gamma\beta} \ S_2 + {\cal O}\left (\beta^{-3/2} \right ) \ .
\label{Seff}
\end{equation}
We have computed first three terms in the above expansion. Here we give 
$S_0$ and $S_1$ 
\begin{equation}
S_0 \ = \ -\frac{1}{2}\sum_{x,n}
\left (R_k(x)-R_k(x+e_n)+i\tilde{j}_k(l) \right )^2 \ ,
\label{Sefflead}
\end{equation}
\begin{equation}
S_1 \ = \ \epsilon^{imk} \ \sum_x W_{im}(x) R_k(x)  \ ,
\label{Seff1}
\end{equation}
\begin{equation}
W_{im}(x) \ = \ p_i(l_1)p_m(l_2)+p_i(l_3)p_m(l_4)  \ , 
{\rm if} \ x \ {\rm is \ even} \ , 
\label{Wimxeven}
\end{equation}
\begin{equation}
W_{im}(x) \ = \ p_i(l_1)p_m(l_4)+p_i(l_3)p_m(l_2)  \ , 
{\rm if} \ x \ {\rm is \ odd} \ .  
\label{Wimxodd}
\end{equation}
We have introduced here the following notations 
$$
p_k(l_1)= R_k(x)-R_k(x+e_1)+i\tilde{j}_k(x,e_1) , 
p_k(l_2)= R_k(x)-R_k(x+e_2)+i\tilde{j}_k(x,e_2) , 
$$
$$
p_k(l_3)= R_k(x-e_1)-R_k(x)+i\tilde{j}_k(x-e_1,e_1) ,
p_k(l_2)= R_k(x-e_2)-R_k(x)+i\tilde{j}_k(x-e_2,e_2) \ .
$$
Due to the term $R^2(x)$ in the measure, summation over $R(x)$ can be 
extended to include point $R(x)=0$. We use Poisson summation formula, 
perfrom change of variables $R(x)\to\sqrt{\gamma\beta}R(x)$ and make 
the following expansion 
\begin{equation}
R_k(x) \ = \ \frac{1}{\sqrt{\gamma\beta}} \ t_k(x)
\sqrt{\gamma\beta R^2(x)-1} \approx t_k(x)R(x) + {\cal O}(\beta^{-1}) \ . 
\label{smclexp}
\end{equation} 
This expansion is valid when $\gamma\beta R^2(x)>1$. We call this expansion 
semiclassical approximation. It allows to define an obvious analog of 
the vortex--spin-wave representation for the PF
\begin{eqnarray}
\label{vortspwrepr}
Z_{\overset{\star}\Lambda}(j,\beta ) &=&  C(\beta ) 
\prod_{x\in\overset{\star}\Lambda} \left [ \sum_{m(x)=-\infty}^{\infty}
\int_{-\infty}^{\infty} \prod_{k=1}^3dR_k(x) \right ]  \\ 
&\times& \exp \left [ S_{eff}(R_k(x)) + 
2i\pi \sqrt{\gamma\beta}\sum_x m(x) 
\left ( \sum_kR^2_k(x) \right )^{1/2} \right ] \ .
\nonumber  
\end{eqnarray}

\subsection{$ISO(2)$-like approximation}

Here we investigate the dual model given by Eq.(\ref{PFLsu2}) with 
the corresponding dual weight (\ref{linkint}). 
Our approach is to replace the matrix elements in the integrand 
of (\ref{linkint}) by the $ISO(2)$-like matrix elements. Justification 
of such replacement will be given in the course of calculations. 
For $SU(2)$ matrix elements we use the Wigner $D$-function parametrized 
by Euler angles
\begin{equation}
D^{mn}_r(\alpha,\omega,\gamma )=e^{-im\alpha -in\gamma}d^{mn}_r(\omega) \ .
\label{Dfunc}
\end{equation}
\noindent
For the sake of simplicity we consider here only the standard action
$H[V]=\sum_nD_{1/2}^{nn}$, i.e. the fundamental character which  in this 
parameterization reads
\begin{equation}
\sum_nD_{1/2}^{nn} = 
2 \cos \frac{\omega}{2} \cos\frac{1}{2}(\alpha + \gamma ) \ .
\label{fundtrace}
\end{equation}
\noindent
The invariant measure on the group takes the form
\begin{equation}
\int  dV = \frac{1}{16\pi^2}\int_0^{4\pi}d\gamma
\int_0^{2\pi}d\alpha \int_0^{\pi}d\omega \sin \omega \ .
\label{invmeas}
\end{equation}
\noindent
Substituting last expressions into (\ref{linkint}) one can exactly integrate
over $\alpha$ and $\gamma$ angles. This gives
\begin{equation}
\Xi_0(l) = \delta_{t_2-t_1}^{m_1-m_2}
\int_0^{\frac{\pi}{2}}d\omega \sin\omega \cos\omega \
I_{2(m_1-t_2)}(2\beta \cos\omega ) \
d^{m_1m_2}_{r_1}(2\omega)d^{t_2t_1}_{r_2}(2\omega ) \ .
\label{xio1int}
\end{equation}
\noindent
To get the asymptotics when $\beta\to\infty$ we use the fact 
(described in the previous subsection) that the link matrix in this region 
performs small fluctuations around unit matrix, namely 
\begin{equation}
\omega\sim {\cal O}(\beta^{-1/2}) \ .
\label{fluct}
\end{equation}
It seems to be unfeasible task to solve saddle-point equation 
for the integrand of (\ref{xio1int}). We therefore take a different 
approach to the problem, and replace this integrand by its asymptotics 
at $\omega\approx 0$ such that 
\begin{equation}
R_i \ \omega \sim {\cal O}(1) \ , \ m_1+m_2 - {\mbox {is arbitrary}} \ , \ 
m_1-m_2 - {\mbox {is fixed}} \ .
\label{condasymp}
\end{equation}
Here $R_i$ is classical angular momentum defined in Eq.(\ref{clangm}). 
The last condition states that the main contribution to the integral comes 
from diagonal components of $SU(2)$ matrix elements, non-diagonal 
contributions can be treated perturbatively. 
To be precise, we adjust the following three approximations:

\begin{enumerate} 

\item
We use the following asymptotics for the
modified Bessel function
\begin{equation}
I_n(x) = \frac{e^x}{\sqrt{2\pi x}} \ \exp (-\frac{1}{2x}n^2)
\left ( 1+ {\cal O}(x^{-1}) \right )
\label{Besselasym}
\end{equation}
\noindent
when $x\to\infty$ and such that $n^2/x\sim {\cal O}(1)$. It leads to
\begin{equation}
I_{2s}(2\beta\cos\omega) = \frac{e^{2\beta}}{\sqrt{4\pi \beta}} \
\exp (-\frac{1}{\beta}s^2 - \beta\sin^2\omega )
\left ( 1+ {\cal O}(\sin^2\omega) \right ) \ .
\label{Bessas1}
\end{equation}
\noindent
Due to the property (\ref{fluct}) the remainder is bounded like
${\cal O}(\beta^{-1})$. 

\item
The integration region over $\omega$ is extended to infinity after 
proper change of variables (this introduces only
exponentially small corrections which can be properly bounded). 

\item
The main approximation concerns the behaviour of matrix elements 
$d^{mn}_r(\omega)$ in the vicinity $\omega\approx 0$. The standard results 
on the asymptotic expansion of $d^{mn}_r(\omega)$ assert that when 
$\omega\to 0$, $r\to\infty$ the $SU(2)$ matrix elements approach
the $ISO(2)$ matrix elements \cite{matrelem,vilenkin}
\begin{equation}
d^{mn}_r(\omega) \ \asymp \  J_{m-n}(r\omega ) \ ,
\label{iso2asympt}
\end{equation}
\noindent
where $J_m(x)$ is the Bessel function. We have found, however that this 
approximation is oversimplified. In particular, it does not lead to the 
correct Gaussian distribution of the dual Boltzmann weight that is clearly 
seen from the effective theory obtained above. Moreover, Eq.(\ref{iso2asympt}) 
states that, e.g. the asymptotics of all diagonal elements is the same. 
To get more precise asymptotics one can use representation of $d$-function 
in terms of the Jacobi function. The latter can be treated with help 
of the Tricomi expansion into degenerate hypergeometric functions as described 
in \cite{bet}, section 10.14. Finally, we use uniform Tailor expansion for 
the degenerate hypergeometric functions \cite{bet}, section 6.13. 
This leads to the following asymptotics     
\begin{equation}
d^{mn}_r(\omega) \ = \  \xi_{mn}  J_k(R \sin \theta \sqrt{b}) 
+ {\cal O}(\sin^2\omega) \ ,
\label{iso2likeasympt}
\end{equation}
\noindent
where all notations are given in (\ref{finnote}). It can be easily checked 
with help of Mathematica that the last equation gives better approximation 
to $d$-function compared to Eq.(\ref{iso2asympt}), especially at large values 
of $m+n$ and is valid, in fact for all values of $r$. 
Moreover, it gives very reasonable approximation in 
rather wide region of $\omega$, except region $\omega\approx\pi/2$.  
In the appendix B we give independent and simpler proof of this asymptotic 
expansion calculating also the first correction ${\cal O}(\sin^2\omega)$ 
to Eq.(\ref{iso2likeasympt}). The final result is given in 
Eq.(\ref{dfuncasympt}). We term this asymptotics as $ISO(2)$-like 
approximation of $SU(2)$ matrix elements. The only (but essential) 
difference from the exact $ISO(2)$ matrix element is the presence of 
$\sin \theta$ in the argument of the Bessel function. 

\end{enumerate}

Combining Eqs.(\ref{Bessas1}), (\ref{iso2likeasympt}) we arrive finally 
at the following asymptotic representation for the one-link integral
\begin{equation}
\Xi_0(l) \ = \ C(\beta ) \ 
\delta^{m_1-m_2}_{t_2-t_1} \ 
\exp \left (-\frac{1}{4\beta} \alpha^2 \right ) B(l)
\left ( 1+ {\cal O}(\beta^{-1}) \right ) \ ,
\label{xioasymp}
\end{equation}
\noindent
where we have used notations
\begin{equation}
C(\beta ) = \frac{e^{2\beta}}{2\beta \sqrt{\pi \beta}} \ ,
\label{CB}
\end{equation}
\noindent
\begin{equation}
\alpha = m_1+m_2-t_1-t_2 \ .
\label{alphdef}
\end{equation}
\noindent
Making change of variables $\sin\omega = y/\sqrt{2\beta}$ in the last integral
and extending the integration region over $y$ to infinity
the last integral can be written as
\begin{equation}
B(l) = \int_0^{\infty}dy \ y \ e^{-\frac{1}{2}y^2} \
J_{m_1-m_2}\left (\frac{R_1\sin\theta_1}{\sqrt{2\beta}} y  \right )
J_{t_2-t_1}\left (\frac{R_2\sin\theta_2}{\sqrt{2\beta}} y  \right ) \ ,
\label{Blink}
\end{equation}
\noindent
where $R$ and $\sin\theta$ are defined in Eq.(\ref{finnote}).
Calculating the last integral we find
\begin{eqnarray}
\Xi_0(l) \ \asymp \ C(\beta) \ \delta_{t_2-t_1}^{m_1-m_2} \ 
\exp \left [ -\frac{1}{4\beta}(R_1^2+R_2^2-2R_1R_2
\cos\theta_1\cos\theta_2) \right ] \nonumber  \\
\times \
I_k\left(\frac{R_1\sin\theta_1 R_2\sin \theta_2}{2\beta} \right) \ ,
\label{xio2das}
\end{eqnarray}
\noindent
where $k=m_1-m_2=t_2-t_1$. Again, we have compared the present asymptotics 
with the exact expression given by Eq.(\ref{xio2dexact}) using Matemathica. 
Suprisingly, if $\beta$ is sufficiently large the asymptotic formula 
gives very reasonable approximation for all configurations we have studied. 
It is interesting to stress that the same asymptotics can be obtained directly 
from (\ref{xio2dexact}), where the CG coefficients are replaced by their 
semiclassical limit by Ponzano-Regge \cite{regge}.  The method presented above 
has however two advantages. First of all, we have computed corrections to 
the leading behaviour (\ref{iso2likeasympt}), therefore 
${\cal{O}}(\beta^{-1})$-corrections to the asymptotic expansion of one-link 
integral can be computed in a straightforward manner. Corrections to 
the Ponzano-Regge formula are in general unknown. Secondly, we find that 
our method is much simpler.

\section{Discussion}

In this article we proposed dual formulation of non-abelian spin models. 
Our approach to the dual transformations is summarized in the beginning 
of the section 2.2. Main formulae of the present paper, 
Eqs.(\ref{PFsundual})-(\ref{2funcdualw}), (\ref{dual2su2})-(\ref{dualBW1}) 
give two versions of the dual formulation. 
To demonstrate the usefulness of our formulations we have 
computed the low-temperature asymptotics of the dual Boltzmann weights 
for two-dimensional model. We reckon that the results for the dual weights 
provide significant simplification of the partition function at large values 
of $\beta$. Nevertheless, even in this case the model is still too 
complicated to be solved exactly. In our next paper \cite{sps} we consider 
some physical applications of the dual representation. In particular, we 
attempt to give a saddle-point solution of the dual model and derive 
effective $1D$ model for the two-point correlation function. Also, we 
calculate continuum limit of the dual representation and solve 
classical continuum equations of motion.  

Some simple consequences which worth noting can be obtained from 
(\ref{vortspwrepr}). Let us perform a shift of variables in the integrand 
of (\ref{vortspwrepr})
\begin{equation}
R_k(x) \ \to \ R_k(x) - \frac{i}{2\gamma\beta}\sum_l 
j_k(l)D_l(x) \ .
\label{shift}
\end{equation}
This shift allows to extract the main contribution from the sources 
to the free energy which emerges from the Gaussian term 
in the effective action, i.e. $S_0$ given by Eq.(\ref{Sefflead})
\begin{equation}
Z_{\overset{\star}\Lambda}(j,\beta ) \ = \  C(\beta )
\exp \left [ \frac{1}{2\gamma\beta}\sum_{ll^{\prime}}
j_k(l)G_{ll^{\prime}}j_k(l^{\prime}) \right ] 
\tilde{Z}(j,\beta ) \ ,
\label{ZjL}
\end{equation}
where $\tilde{Z}(j,\beta )$ is a remaining part of the PF. 
Here $D_l(x)$ and $G_{ll^{\prime}}$ are link Green functions introduced 
in \cite{su2}. In \cite{sps} we show that the following sources 
$$
j_k(l) \ = \ -2i\sqrt{j(j+1)}h_k \ , \ l\in C_{xy} \ , \ 
h^2_k = 1  
$$
and $j_k(l)=0$, otherwise, can be related to the correlation function 
in the representation $j$. Substituting these sources into (\ref{ZjL}) 
one finds after simple analysis 
\begin{equation}
\Gamma_j(R) \ = \ 
Z_{\overset{\star}\Lambda}(j,\beta )/Z_{\overset{\star}\Lambda}(0,\beta ) 
\ = \  \exp \left [ -\frac{2}{\gamma\beta}j(j+1)D(R) + 
{\cal O} (\beta^{-3/2}) \right ] \ ,
\label{Gjlead}
\end{equation}
where $D(R)={\cal O}(\ln R)$, for $R\to\infty$. By no means we claim that 
this result establishes power-like decay of the correlation function: 
in fact, we do not know the $R$-dependence of the remainder. Therefore, 
at the moment we can only claim that the shift (\ref{shift}) allows to extract 
the main perturbative contribution. We would like to stress that the shift 
(\ref{shift}) is precisely the same as the one used in \cite{rigbkt}. 
Unfortunately, we do not know if the methods of \cite{rigbkt} can be 
applied to the partition function (\ref{ZjL}) to bound the remainder:
the essential difference from the $XY$ model is that even in the absence 
of vortices both full and effective dual actions are complex. 
Nevertheless, we think that such possibility deserves further investigations.

\begin{appendix}

\section{Asymptotics of $\ln\Phi (\xi ;V)$}

In this appendix we calculate the asymptotic expansion of the function 
$\ln\Phi (\xi ;V)$, where 
$\Phi (\xi ;V)\equiv\Phi (r,t,\omega )$ is given by 
Eq.(\ref{Phidef})
\begin{equation} 
\Phi (r,t,\omega ) \ = \ \sum_{\lambda =0}^{2r} (-i)^\lambda \ 
\frac{2\lambda + 1}{2 r + 1} \ P_{\lambda}(t) \ 
\chi_{\lambda}^r \left ( \omega \right ) \ .
\label{Phidef1}
\end{equation} 
Here, $P_{\lambda}(t)$ is the Legendre polynomial and $t$ is given in 
Eq.(\ref{combangle}). $SU(2)$ matrix is taken in the parameterization 
(\ref{Vrmn}), (\ref{su2matr}). Also, we define and compute the semiclassical 
limit of $\Phi (r,t,\omega )$. To this end we  introduce classical angular 
momentum as 
\begin{equation}
R \ = \ 2r+1 \ .
\label{clangm}
\end{equation}
We are interested in the expansion of $\Phi (\xi ;V)$ when $\omega\approx 0$ 
uniformly valid in all variables $\xi$. 
Because $\varphi$ and $\alpha$ are compact variables 
the problem of uniformity concerns mainly the value of the variable $r$. 
We face two situations: 1) $r$ is fixed and 2) $r\to\infty$.  In the second case 
we look for the asymptotic expansion such that $R\omega\sim{\cal{O}}(1)$. 

As the first step we express the generalized characters entering
$\Phi (\xi ;V)$ through the associated Legendre functions of the first kind
taken on the cut $x\in [-1,1]$ 
\begin{equation}
\chi^r_{\lambda} (\omega ) = \sqrt{\frac{\pi}{2}}
\left [ \frac{(2 r+1)(2 r+1+\lambda)!}{(2r-\lambda)! \sin\frac{\omega}{2}}
\right ]^{1/2}
{\rm P}^{-\lambda-1/2}_{2 r+1/2} (\cos\frac{\omega}{2}) \ .
\label{chiLeg}
\end{equation}
\noindent
When $r$ is fixed, we construct Tailor expansion of $\Phi (\xi ;V)$. It is 
more convenient to expand in powers of $y=\sin\frac{\omega}{2}$ rather than 
in powers of $\omega$. From the representation (\ref{chiLeg}) it follows that 
\begin{equation}
\chi^r_{\lambda} (\omega ) = \frac{\sqrt{\pi}}{2\Gamma (\lambda +3/2)}
\left [ \frac{R(R+\lambda)!}{(R-\lambda-1)!} \right ]^{1/2} 
\left ( \frac{y}{2}  \right )^{\lambda}
\left [ 1 - \frac{R^2-(\lambda +1)^2}{2\lambda +3}\frac{y^2}{2} + 
{\cal{O}}(y^4) \right ] \ .
\label{chiexp}
\end{equation}
\noindent
Substituting last expression into Eq.(\ref{Phidef1}), it is straightforward 
to find the following expansion 
\begin{equation} 
\ln \Phi (r,t,\omega ) \ = \ -iyF_1(R,t)+\frac{1}{6}y^2F_2(R,t) + 
\frac{i}{6}y^3F_3(R,t) + {\cal{O}}(y^4) \ ,
\label{Phiexp}
\end{equation} 
where  the coefficients of the expansion $F_i(R,t)$ are 
\begin{equation} 
F_1(R,t) \ = \ t \sqrt{R^2-1} \ ,
\label{F1}
\end{equation}  
\begin{equation} 
F_2(R,t) \ = \ (3t^2-1) \left [ R^2-1 - \sqrt{R^2-1}\sqrt{R^2-4}\right ] \ ,
\label{F2}
\end{equation} 
\begin{eqnarray} 
F_3(R,t)  &=&  t \sqrt{R^2-1} \{ t^2 \left [ \sqrt{R^2-4}\sqrt{R^2-9}- 
\sqrt{R^2-1}\sqrt{R^2-4}+2(R^2-1) \right ] \nonumber  \\
&+& \frac{3}{5}(R^2-4)-R^2+1 
- \frac{3}{5}\sqrt{R^2-4}\sqrt{R^2-9} + \sqrt{R^2-1}\sqrt{R^2-4} \}\ .
\label{F3}
\end{eqnarray} 
When $R\gg 1$ we expand coefficients $F_i$ in powers of $R$. 
This defines the semiclassical expansion of $\Phi (\xi ;V)$ 
when $\omega\to 0$, $R\to\infty$ such that $R\omega\sim{\cal{O}}(1)$  
\begin{equation} 
\ln \Phi (r,t,\omega ) \ = \ -iRty + \frac{i}{2R}ty + 
\frac{1}{4}y^2 \left ( 3t^2-1-\frac{2i}{3}Rt^3y \right ) 
+ {\cal{O}}(y^4) \ .
\label{Phisemiclexp}
\end{equation}

It would seem that the bound ${\cal{O}}(y^4)$ in (\ref{Phiexp}) and, therefore in 
(\ref{Phisemiclexp}) does not hold uniformly in $R$. It turns out however 
that the bound is correct both for fixed $r$ and for $r\to\infty$.
To prove this we need an expansion for the associated Legendre
functions at large $r$ uniformly valid in the neighbourhood
of the point $\omega =0$. Such an expansion is given by
MacDonald's formula \cite{bet}, section 3.5
\begin{eqnarray}
\label{Legasymp}
{\rm P}^{-\mu}_{\nu}(\cos\omega ) &=&
\left [ (\nu +\frac{1}{2})\cos\frac{\omega}{2} \right ]^{-\mu} \\   
&\times& \left \{ J_{\mu}(\alpha ) +
\sin^2\frac{\omega}{2}\left [ \frac{\alpha}{6}J_{\mu +3}(\alpha ) -
J_{\mu +2}(\alpha ) + \frac{1}{2\alpha}J_{\mu +1}(\alpha ) \right ] +
{\cal O}(\omega^4) \right \} \ , \nonumber 
\end{eqnarray}
\noindent
where $\alpha = (2\nu +1)\sin\frac{\omega}{2}$. 
The ratio of factorials in (\ref{chiLeg})
is expanded as 
\begin{equation}
\left [ \frac{(R+\lambda)!}{(R-\lambda -1)!}\right ]^{1/2} =
R^{\lambda +1/2}\left [1-\frac{\lambda (\lambda^2+\frac{3}{2}\lambda
+ \frac{1}{2})}{6R^2} + {\cal O}(R^{-4}) \right ] \ .
\label{ratiofact}
\end{equation}
With these results, and using recursion relations for the Bessel functions
to reduce order to $\lambda +\frac{1}{2}$ we obtain
the representation for the generalized character of the form
\begin{equation}
\chi^r_{\lambda} (\omega) = \sqrt{\frac{\pi}{2}}
\left [ \frac{R}{2\sin\frac{\omega}{4}} \right ]^{1/2} \
B(R,\lambda ;\omega,\partial_h)\mid_{h=1} \
J_{\lambda +\frac{1}{2}}(\alpha h) \ .
\label{chiasymp}
\end{equation}
\noindent
The function $B(R,\lambda ;\omega )$ has the following asymptotic expansion
\begin{equation}
B(R,\lambda ;\omega ,\partial_h) =
1+\frac{\sin^2\frac{\omega}{4}}{3}\left [\frac{7}{4}
-\left ( \frac{2\lambda}{\alpha^2}(\lambda +1)
-\frac{1}{2} \right )
\frac{\partial}{\partial h}\right ] + {\cal O}
\left ( \sin^4\frac{\omega}{4} \right ) \ ,
\label{Basymp}
\end{equation}
\noindent
where $\alpha = 2R\sin\frac{\omega}{4}$. Substituting last expressions into 
Eq.(\ref{Phidef1}) we can easily calculate all the sums and derivatives. 
This leads to 
\begin{eqnarray} 
\label{PhiRwexp}
\ln \Phi (r,t,\omega ) &=& -2iRt\sin\frac{\omega}{4} + 
\frac{i}{R}t\sin\frac{\omega}{4}   \\
&+& \sin^2\frac{\omega}{4} \left ( 3t^2-1-\frac{4i}{3}Rt^3
\sin\frac{\omega}{4} + iRt\sin\frac{\omega}{4} \right ) 
+ {\cal{O}} \left ( \sin^4\frac{\omega}{4} \right ) \ . \nonumber  
\end{eqnarray}
Re-expanding last formula in powers of $\sin\frac{\omega}{2}$ one sees 
that Eq.(\ref{PhiRwexp}) coincides with Eq.(\ref{Phisemiclexp}). Due to 
the bound ${\cal O}(\omega^4)$ in (\ref{Legasymp}), the same 
bound holds in (\ref{PhiRwexp}) and, consequently in (\ref{Phisemiclexp}) 
and in (\ref{Phiexp}). We thus conclude that the Tailor expansion 
(\ref{Phiexp}) does provide expansion at small $\omega$ which is valid 
both for fixed values of $r$ and for $r\to\infty$.

\section{Asymptotics of $d^{mn}_r(\omega)$}

Here we compute the asymptotic expansion of the $SU(2)$ matrix elements
$d^{mn}_r(\omega)$ in the classical region $R=(2r+1)\gg 1$ uniformly
valid in the vicinity of the point $\omega =0$ for all allowed values
of $m$ and $n$, $m-n$ is held fixed. 

To get such an asymptotics we first present $d$-function in terms of hypergeometric
function $F \equiv {}_2F_1$
\begin{eqnarray}
d^{mn}_r(\omega)=\frac{\xi_{mn}}{k!}\left[\frac{(s+k+p)!(s+k)!}{s!(s+p)!}
\right]^{\frac{1}{2}}
\left(\sin \frac{\omega}{2} \right)^k \left(\cos \frac{\omega}{2} \right)^{-p}
\nonumber \\
\times F(s+k+1,-s-p;k+1;\sin^2 \frac{\omega}{2}) \ ,
\label{dasF}
\end{eqnarray}
\noindent
where $\xi_{mn}=1$ if $n\geq m$, $\xi_{mn}=-1$ otherwise, and
\begin{equation}
k=|m-n|, \ p=|m+n|, \ s=r-\frac{1}{2}(k+p) \ .
\label{notat}
\end{equation}
\noindent
As is seen from the arguments of the hypergeometric function the infinite
series in $F$ terminates so that right-hand side of (\ref{dasF}) is
polynomial in $\sin^2\frac{\omega}{2}$
\begin{eqnarray}
d^{mn}_r(\omega) = \xi_{mn} \left [ A_k(x)A_k(y) \right ]^{\frac{1}{2}}
\left(\sin \frac{\omega}{2} \right)^k \left(\cos \frac{\omega}{2} \right)^{-p}
\nonumber \\
\times \sum_{l=0}^{r+\frac{1}{2}(p-k)}(-1)^l
\frac{(\sin^2\frac{\omega}{2})^l}{\Gamma (k+1+l)l!}{\cal F}_l(x,y) \ ,
\label{dseries}
\end{eqnarray}
\noindent
where we have introduced the following notations
\begin{equation}
A_k(x) = \frac{\Gamma (x-\frac{1}{2}k+\frac{1}{2})}
{\Gamma (x+\frac{1}{2}k+\frac{1}{2})} \ ,
\label{Akx}
\end{equation}
\noindent
\begin{equation}
{\cal F}_l(x,y) = \frac{\Gamma (x+\frac{1}{2}k+\frac{1}{2})}
{\Gamma (x-\frac{1}{2}k+\frac{1}{2}-l)}
\ \frac{\Gamma (y+\frac{1}{2}k+\frac{1}{2}+l)}
{\Gamma (y-\frac{1}{2}k+\frac{1}{2})}  \ ,
\label{Flxy}
\end{equation}
\noindent
\begin{equation}
x = r+\frac{1}{2}(1+p), \ y = r+\frac{1}{2}(1-p) \ .
\label{xydef}
\end{equation}
\noindent
The second step consists in expanding the ratio of  Gamma functions. This can be
done with help of the following formula
\begin{equation}
\frac{\Gamma(x+a)}{\Gamma(x+b)}=x^{a+b} \left(
\sum_{s=0}^{N-1}\frac{(-1)^s}{s!x^s}
B_s^{(a-b+1)}(a)(b-a)_s+{\cal O}(x^{-N}) \right) \ ,
\label{Gratio}
\end{equation}
\noindent
where $B_s^{(y)}(x)$ are the generalized Bernoulli polynomials.
Important point concerns the large expansion parameter we use.
We take not merely classical angular momentum $r+\frac{1}{2}$
but rather quantities $x$ and $y$ defined above. Such a choice
gives more accurate asymptotics valid in a wider region of parameters.
Then, in the case of quantity ${\cal F}_l(x,y)$ the series in (\ref{Gratio})
terminates because $a-b=k+l$ and representation (\ref{Gratio}) becomes exact.
It leads to
\begin{eqnarray}
{\cal F}_l(x,y) = (xy)^{l+k}\sum_{s_1,s_2=0}^{l+k} \frac{1}{x^{s_1}y^{s_2}s_1!s_2!}
\frac{[(l+k)!]^2}{(l+k-s_1)!(l+k-s_2)!} \nonumber  \\
\times B_{s_1}^{(k+l+1)}(\frac{1}{2}(k+1)) B_{s_2}^{(k+l+1)}(\frac{1}{2}(k+1)+l) \
\label{Flxyexp}
\end{eqnarray}
\noindent
what is essentially the desired expansion at large $x$ and $y$.
It follows from the last representation that
\begin{eqnarray}
{\cal F}_l(x,y) = (xy)^{l+k} [ 1+\frac{1}{2}l(l+k)(\frac{1}{y}-\frac{1}{x}) -
\frac{l^2(l+k)^2}{4xy}  \nonumber  \\
+ \frac{1}{24}(l+k)(l+k-1)(3l^2-l-k-1)(\frac{1}{x^2} +
\frac{1}{y^2}) + {\cal O}(x^{-3},y^{-3}) ] \ .
\label{Flxyasym}
\end{eqnarray}
\noindent
For $A_k(x)$ following the same procedure one finds
\begin{equation}
A_k(x) = x^{-k}\left [1+\frac{1}{24x^2}k(k^2-1) + {\cal O}(x^{-4}) \right ] \ .
\label{Akxasym}
\end{equation}
\noindent
Substituting last expressions into (\ref{dseries}) we get after some algebra
\begin{eqnarray}
d^{mn}_r(\omega)=\xi_{mn}(\cos \frac{\omega}{2})^{-p}
\sum_{l=0}^{\infty} \frac{(-1)^l}{(l+k)!l!} \left(
\frac{t}{2}\right)^{k+2l}   \nonumber  \\
\{ 1+\frac{1}{2} l(l+k) \left( \frac{1}{y}-\frac{1}{x} \right)
-\frac{l^2(l+k)^2}{4xy}+\frac{k(k^2-1)}{48} \left( \frac{1}{x^2}
-\frac{1}{y^2}\right)+ \\
\frac{1}{24} (l+k)(l+k-1)(3l^2-l-k-1) \left(
\frac{1}{x^2} -\frac{1}{y^2} \right) \} \ .  \nonumber
\label{dexpB}
\end{eqnarray}
\noindent
\noindent
Here we have extended summation over $l$ to infinity since this introduces
corrections of the order ${\cal O}(\omega^{2r})$ or less.
Recalling now the series representation for the Bessel function
\begin{equation}
J_k (t) = \sum_{l=0}^{\infty} \frac{(-1)^l}{(l+k)!l!}\left(\frac{t}{2}
\right)^{k+2l}
\label{Bessel}
\end{equation}
\noindent
we can easily sum up all series in the last formula. Finally, we arrive at the
following asymptotic expansion for $d$-function
\begin{eqnarray}
d^{mn}_r(\omega)= \xi_{mn} \{ J_k(t)+\frac{b}{4} [ J_k(t)+\frac{t}{3\sin^2\theta}
(1-2\cos^2\theta)(J_{k-1}(t)-  \nonumber  \\
J_{k+1}(t)) - \frac{1+\cos^2 \theta}{6\sin^2\theta}
\left[(k+1)J_{k-2}(t)-(k-1)J_{k+2}(t) \right]]+
O (\sin^4\frac{\omega}{2}) \} \ .
\label{dfuncasympt}
\end{eqnarray}
\noindent
We have introduced here the following notations:
\begin{equation}
R=2r+1, \ \cos \theta = \frac{p}{R}, \ b=\sin^2 \frac{\omega}{2}, \
t=R \sin \theta \sqrt{b} \ .
\label{finnote}
\end{equation}
\noindent

\end{appendix}


\begin{thebibliography}{99}


%
\bibitem{savit} R.~Savit, Phys.Rev.Lett. 39 (1977) 55;
Rev.Mod.Phys. 52 (1980) 453.
%
\bibitem{rigbkt} J.~Fr\"ohlich, T.~Spencer, Commun.Math.Phys.
81 (1981) 527.
%
\bibitem{pfeiffer} H.~Pfeiffer, J.Math.Phys. 44 (2003) 2891.
%
\bibitem{linkrepr} G.~Batrouni, M.B.~Halpern, Phys.Rev. D30 (1984) 1775.
%
\bibitem{su2} O.~Borisenko, V.~Kushnir, A.~Velytsky,
Phys.Rev. D62 (2000) 025013.
%
\bibitem{2ddual} O.~Borisenko, V.~Kushnir, Low-temperature behaviour of $2D$ 
lattice $SU(2)$ spin model, Proc. of NATO Workshop
"Integrable structures of exactly solvable two-dimensional models of quantum
field theory", Ed. by S.~Pakuliak, and G.~von Gehlen,
Kluwer Academic Publishers, 2001, 55.
%
\bibitem{seiler} A.~Patrascioiu, E.~Seiler, J.Statist.Phys. 69 (1992) 573;
J.Statist.Phys. 106 (2002) 811-826.
%
\bibitem{mgsu2} O.~Borisenko, P.~Skala, Phys.Rev. D62 (2000) 014502.
%
\bibitem{dualu1}  T.~Banks, J.~Kogut, R.~Myerson, Nucl.Phys. B121 (1977) 493.
%
\bibitem{3du1lgt} M.~G\"{o}pfert, G.~Mack, Commun.Math.Phys. 81 (1981) 97; 82
(1982) 545.
%
\bibitem{matrelem} D.A.~Varshalovich, A.N.~Moskalev, V.K.~Khersonskii,
Quantum theory of angular momentum, World Scientific Publishing Co.Pte.Ltd.,
Singapore-New Jersey-Hong Kong, 1988.
%
\bibitem{klymik} N.Ja.~Vilenkin, A.U.~Klimyk, Representation of Lie groups 
and special functions,  Kluwer Academic Publishers, Dordrecht-Boston-London, 
Vol.316, 1995.  
%
\bibitem{contest} J.~Bricmont and J.-R.~Fontaine, J.Stat.Phys. 26 (1981) 745.
%
\bibitem{vilenkin} N.Ja.~Vilenkin, Special functions and theory of group 
representations, Nauka, Moscow, 1991. 
%
\bibitem{bet} H.~Bateman, A.~Erdelyi, Higher Transcendental
Functions, New York-Toronto-London, 1953; Russian Edition, 
Nauka, Moscow, Vol.1,2, 1973. 
%
\bibitem{regge} G.~Ponzano, T.~Regge, Semiclassical limit of Racah 
coefficients, in Spectroscopic and group theoretical methods in physics, 
North-Holland Publ. Co., Amsterdam, 1968, 1; K.~Schulten, R.G.~Gordon, 
Journal of Math.Phys., V16 (1975), 1971.
%
\bibitem{sps} O.~Borisenko, V.~Kushnir, Saddle-point solution of the dual 
of $2D$ $SU(2)$ chiral model, in preparation. 
%

\end{thebibliography}
\end{document}